\documentclass{llncs}

\usepackage{booktabs} 

\usepackage{graphicx}
\usepackage{cellspace}
\usepackage{listings}
\usepackage{wrapfig,picins}
\newcommand\cincludegraphics[2][]{\raisebox{-0.3\height}{\includegraphics[#1]{#2}}}

\usepackage{amsmath}
\usepackage{graphicx}
\usepackage{color}
\usepackage{url}
\usepackage{cite}
\usepackage{booktabs} 
\usepackage{subfigure}

\usepackage{amsmath}

\DeclareGraphicsExtensions{.pdf,.png,.jpg}

\usepackage[english]{babel}

\begin{document}

\frontmatter          
\pagestyle{empty}  

\title{An immersive Open Source environment using Godot}
%
%
%
 \author{Francesca Santucci\inst{1} \and Federico Frenguelli\inst{1} \and Alessandro De Angelis\inst{1} \newline Ilaria Cuccaro\inst{3}\newline Damiano Perri\inst{2} $^{ORCID: 0000-0001-6815-6659}$\newline Marco Simonetti\inst{2} $^{ORCID: 0000-0003-2923-5519}$
 }
\institute{Evonove SrL, Perugia, Italy\and
University of Florence, Dept. of Mathematics and Computer Science, Florence, Italy \and University of Perugia, Dept. of Mathematics and Computer Science, Perugia, Italy
}

\maketitle

\begin{abstract}
We present a sample implementation of a Virtual and Augmented Reality immersive environment based on Free and Libre Open Source Hardware and Software and the HTC Vive system, used to enhance the immersive experience of the user and to track her/his movements.

The sense of immersion has increased and stimulated using a footplate and a Tibetan bridge, connected to the virtual world as Augmented Reality applications and implemented through an Arduino board, thereby adopting a low cost, open source hardware and software approach.

The proposed architecture is  relatively affordable from the cost point of view, easy to implement, configure and adapt to different contexts. 
It can be of great help for organizing laboratory classes for young students to afford the implementation of virtual worlds and Augmented Reality applications. 
\end{abstract}


\section{Introduction}
The importance of virtual reality grows with the spread of digital technologies in various areas of life.  The year 2020 will remain sadly etched in our minds and memories for having deprived us, in various forms, of cherished affections and our lifestyles. In a few months we have seen the use of technology grow disproportionately.

Virtual reality can allow us to explore new forms of communication and distance education, becoming one of the most promising areas for the future, allowing new forms of teaching, capable of bridging the physical distance and giving a new strength to online education. 

In a virtual world, 
the system follows and simulates our movements through images and sounds that give us the illusion that we are moving and acting in a synthetic world. These sensations become much deeper and more suggestive when we are able to visit the world with immersive devices, like the HTC Vive system, consisting of a base station, hand motion controllers and a Head Mounted Display.

 The emotional involvement during a virtual reality experience depends on the quality of the stimulus that involves the player. Suppose that a player is moved by a series of bodies that will change his point of view, so that the player could have the feeling to be moved. To create a more immersive experience, we may add some stimuli (like a vibration system) that are able to move the player in reality.
 
The purpose of the project  is to develop, with \textit{open source} tools and software, an interactive virtual reality experience that allows the player to do a charming and attractive experience, using objects mapped in the virtual world with which the player can interact. In our case we implemented a footplate that is moved by an Arduino system and a Tibetan bridge oscillating along the x-axis.
The implemented system represents a low cost, open source and easily reproducible case study, that allows young students to experience the immersive learning.

The paper is organized  as follows: in section \ref{relwork} a review of publication related the paper keywords is carried out, in section \ref{arch} the architecture of the system is described, in section \ref{virtualworld} the virtual world implemented on Godot and in Blender is described. The section \ref{conclusions} presents some conclusions and a description of future works.

\section{Related works \label{relwork}}
Since the construction of the first virtual reality viewer in the 1970s, considerable progress has been made year after year.
The improvements we are seeing are both hardware and software.
The cost of the viewers is decreasing and many new applications are being developed.
The sectors where virtual reality finds profitable application are: entertainment, tourism, manufacturing industry, e-commerce, medicine, teaching\cite{vr2018Casiduso}, and troubleshooting (for example in companies).

In the medical field, it is possible to help patients suffering from brain trauma through tele-rehabilitation.
An example is in the work described in \cite{vrrehabilitation} where a Virtual Reality application based on X3D is presented that allows to stimulate the patient's brain with simple but concrete exercises, such as cross a street at a pedestrian crossing with traffic lights or calling and taking an elevator.

Thanks to modern information technology it is possible for a therapist to dynamically control the patient's status via remote connection, analyze the outcome of practices,  and to program new types of exercises. Tele-rehabilitation increase the demand of calculations to be carried out and of data stored on a central repository.
To this purpose cloud systems integrated with high performance systems are increasingly important solutions to provide the necessary computing resources.
This approach makes it possible to optimise costs and easily afford future development and upgrade plans in relation to the number of users currently using the system and estimating how many will use it in the near future.\cite{VellaFGCS}

Teaching is an area where it is possible to use virtual reality (VR) technologies in a productive manner as students adapt and assimilate the explained notions more quickly.
This is explained by the increased brain stimulation they receive when immersed in a virtual world, which is a completely different type of experience respect to simply read a text in a book.
An example can be found in \cite{Hussein2015TheBO} where the quality of the study of the planets of the solar system improves the use of Virtual Reality.

The use of input devices that integrate the user's movements to the scenarios presented and that allow an increasingly fluid navigation within the network, are of particular importance for the purpose of a deep learning.\cite{Gervasi2009}
Furthermore, interconnected environments where huge quantities of personal information flows continuously require the integration of high security protocols \cite{VellaGPU} and suitably tested performing hardware\cite{VellaNGT12}.
For the purpose of secure access to user accounts, the integration with specific devices and deep learning systems for the recognition of facial features\cite{Riganelli2017} or passwords recognized through the movement of the lips\cite{lipdetection} is very promising.

In order to make the study environment more likely to real spaces and make it more informative, photos taken from ImageNet\footnote{ImageNet is an image database organized according to the WordNet hierarchy (currently only the nouns), in which each node of the hierarchy is depicted by hundreds and thousands of images. it can be found at the URL: \url{http://www.image-net.org}} have been inserted together with rendered images: through the metadata inserted in it\cite{Krizhevsky2012}, the user also experiences an Augmented Reality environment, which allows him to greatly expand his knowledge.

\section{The Architecture of the system\label{arch}}
The project consists in a virtual reality experience in which the player can move and interact with the objects available in the environment.

\begin{figure}
\centering
\cincludegraphics[width=0.4\textwidth]{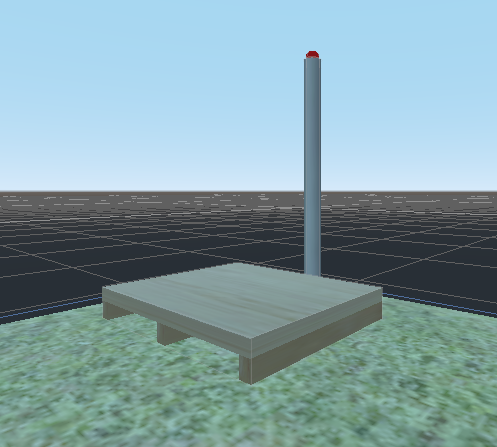}
\caption{The footplate and the button in the initial scene of the virtual environment.\label{fig:footplate}}
\end{figure}

The project was designed thinking about of the real available space. For this reason the experience consists in a circular path: the player will start the game at one edge of the real room, he has to reach a footplate (that is placed both in the virtual and in the real worlds), shown in figure \ref{fig:footplate}, get on it and press a button in the virtual world that will activate its movement. When the footplate is activated, in real world it starts to vibrate while, in virtual world, the player is moved to the upper floor. 
Upstairs there is a path where the player can move and walk until it reaches the bridge. 
The bridge is able to detect the collision with the player so when he starts to walk over it some forces are applied to the bridge to simulate the behavior of a suspended bridge in the vacuum.
The path to be followed by the player is circular in the real world and developed upwards in the virtual world. This technique is designed to solve room spacing's issue and to give the player the feeling of walk a long way.

It is fascinating how the virtual reality can influence the player sensation in the same way of real world. Even if the player knows that all system is virtual and not real, when he has to jump in a vacuum he fells scared and does not want to jump.

\begin{figure}
\centering
\cincludegraphics[width=0.4\textwidth]{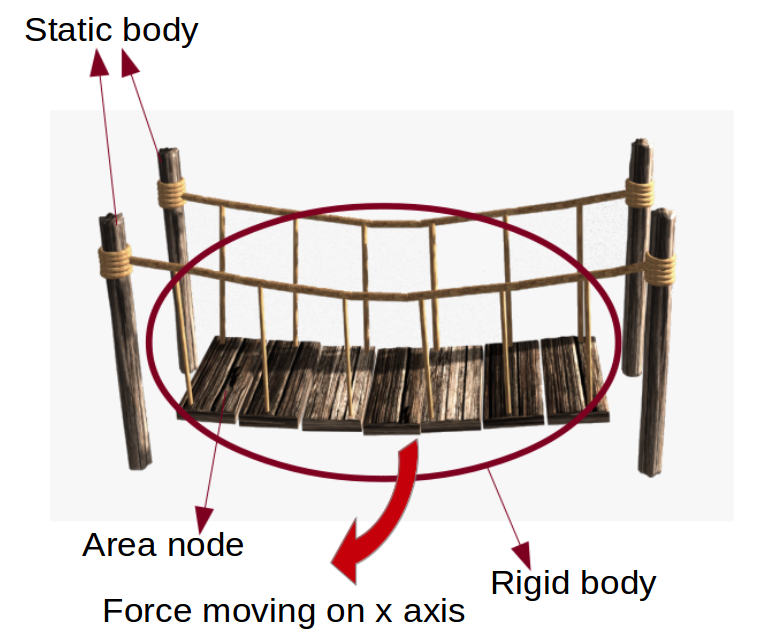}
\caption{The Tibetan bridge.\label{fig:Tibetanbridge}}
\end{figure}

The aim that our project would reach is to mislead the player's feelings. For this reason we made a path developed upwards in which there is a suspended Tibetan bridge, shown in figure \ref{fig:Tibetanbridge}.
Furthermore we add a wooden footplate that is present both in virtual and real world and its position is mapped in the same point. In this way when the player reaches the footplate in virtual world he will reach it in real world too.

\begin{figure}
\centering
\cincludegraphics[width=0.4\textwidth]{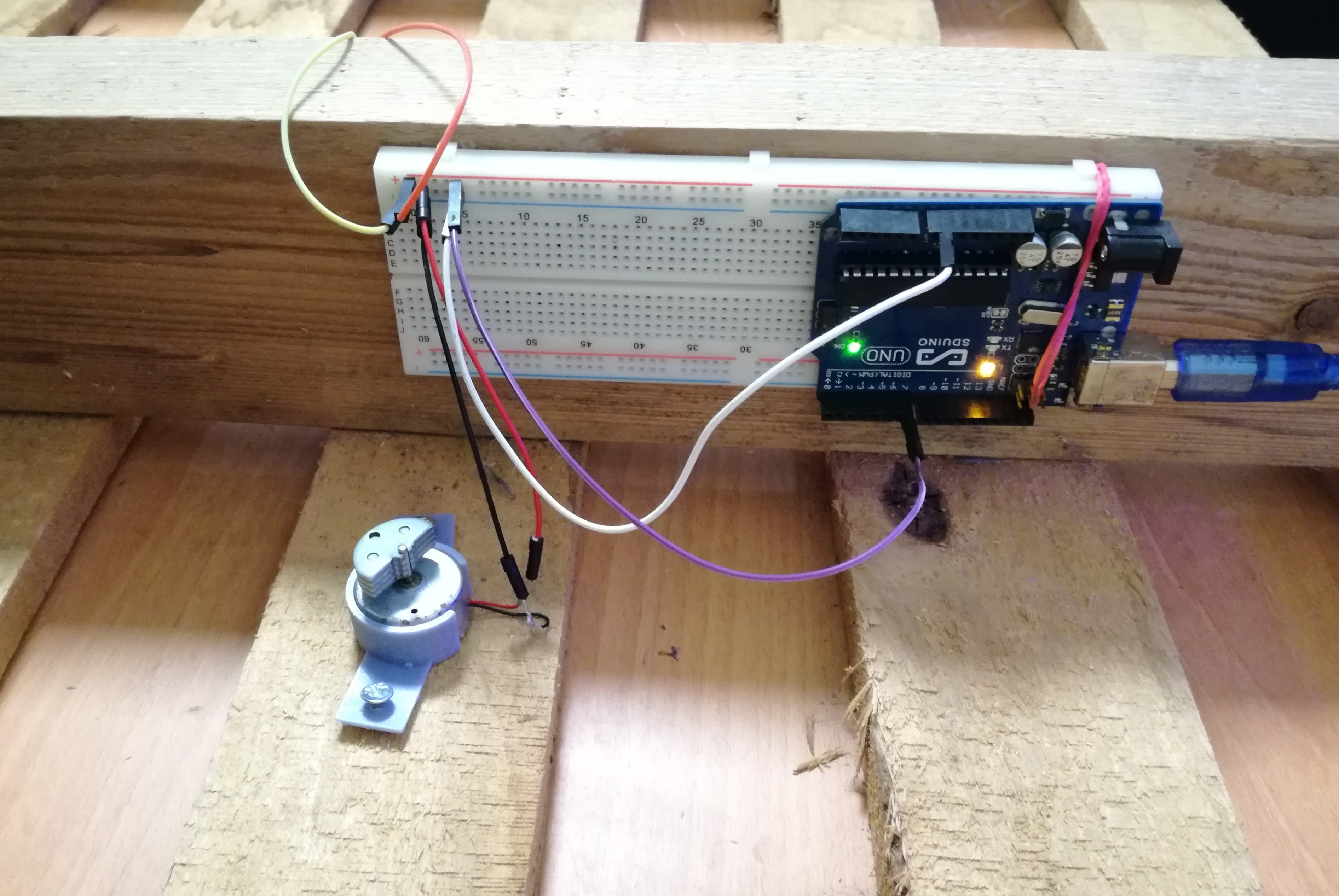}
\caption{The Arduino board installed on the real footplate.\label{fig:arduino}}
\end{figure}

Under the footplate we added some vibration motors, shown in figure  \ref{fig:arduino}, used to mislead the perception of the player. When the player goes on it and presses the button in the virtual world, the motors will be activated. In this way in the real world the player is affected by some vibrations that give him some feelings of instability while in the virtual world he is moved upstairs, ready to start exploring the virtual world.

The development of software for virtual reality requires specialized software and hardware. In fact, virtual reality requires a high frame rate to be able to move smoothly and naturally within a virtual space. This is a core requirement not only for experiencing an enjoyable 3D world, but also in order to prevent the so called “\textit{simulation sickness}”, which might cause dizziness, headaches or even nausea. 

First of all, a headset is needed. Our project was developed to work with \textbf{HTC Vive}. This is a virtual reality headset developed by \textbf{HTC} and \textbf{Valve}. 
The headset uses \textit{room scale tracking technology}, allowing the user to move in 3D space and use motion-tracked handheld controllers to interact with the environment. 
The kit that HTC Vive offers is composed by different components. We use the following:
\begin{itemize}
    \item \textit{Headset:} is a device that uses two OLED panels, one per eye, each having a display resolution of 1080x1200 (2160x1200 combined pixels).
    It has a refresh rate of 90 Hz and 110 degree field of view. 
    The device includes a front-facing camera that allows the user to observe their surroundings without removing their headset. The software can also use the camera to identify any static or moving objects in the room; this functionality can be used as part of \textit{Chaperone} safety system which will automatically display a virtual wall or a feed from the camera to safely guide users from obstacles or real-world walls.
    \item \textit{Controllers:} represent the hands of the player. They have multiple input methods including a track pad, grip buttons, and a dual-stage trigger and a use per change of about 6 hours.
    \item \textit{Base stations:} Also known as the Lighthouse tracking system are two black boxes that create a 360 degree virtual space. The base stations emit timed infrared pulses at 60 pulses per second that are then picked up by the headset and controllers with sub-millimeter precision.
    \item \textit{Trackers:} Are some motion tracking accessory. They are designed to be attached to physical accessories and controllers, so that they can be tracked via the Lighthouse system. In our project we attach trackers to the player’s feet. In this way we do not have to implement the teleport system. \cite{HTC_Vive}
\end{itemize}
The minimum system requirements that the computer must have to work with HTC Vive are at least:
\begin{itemize}
    \item the following \textbf{hardware}:
       processor Intel Core i5-4590 or AMD FX 8350;
       GPU NVIDIA GeForce GTX 970 or AMD Readeon RX 290; memory 4 GB RAM; 
       video output HDMI 1.4, DisplayPort 1.2;
        1 x USB 2.0
    \item and the following \textbf{software}:
        Operating System: Windows 7 SP1, Windows 8.1 or later, Windows 10. \cite{SystemRequirements};
\end{itemize} 

To create a virtual reality experience many software can be used, like Unity 3D, Unreal Engine and Godot. Each software is a game engine and the main difference between them is that Godot is \textit{open source} and does not impose any rules or payments when the game is published.

Probably, to develop a virtual reality experience Unreal Engine and Unity offer a better support and stability, but for the purpose of our project we decided to use Godot. 
Godot is a free open source project developed by a community of volunteers. It provides a huge set of common tools and a fully integrated game development environment.

The main component of Godot are the \textit{Nodes} that can perform a variety of specialized functions. Each node has a name and some editable properties. It can receive a callback to process every frame, can be extended and can be added to another node as child. 
A group of nodes organized hierarchically composed a \textit{Scene}. Basically in Godot, running a game means running a scene. A project can contain several scenes, but for start the game, one of them must be selected as the main scene.

The main goal that comes with instancing scenes is that it works as an excellent design language. When making games with Godot the recommended approach is to think about the scenes in a natural way: imagine all the visible elements in your game, write down them in a diagram and start creating a scene for each of them. This approach dismiss the most common design pattern (like Model-View-Controller) used in another game engine.

To have the ability to detect the collisions, Godot offers a \textit{collision system} that allows to build up complex interactions between a variety of objects that compose the game. Each one of these objects could be of {four} different kinds of physics bodies. Each one reacts in different way to the Godot’s physics engine and for this reason it is important to understand what kind of bodies we need.  
To be able to detect and respond to collisions a \textit{physics body} has to be associated to a \textit{collision shape} that defines the object collision bounds. In this way it is able to detect the contact with other objects.

To interact with headset components and setup a virtual environment a new architecture was introduced in Godot called \textit{ARVRServer}. On top of this architecture, specific implementations are available as interfaces, most of which are plugins based on \textit{GDNative}. GDNative offers a way to integrate features and functions defined inside some external libraries or external modules that must have a C-compatible binary interface. Otherwise these modules or libraries cannot be used inside Godot.
Through GDNative we can register symbols, functions names and structures of some dynamic libraries and load them at runtime. These symbols, functions and structure are made available in Godot through a Native Script.
Once ARVRServer architecture is added in game project, every time Godot starts, each available interface will make itself known to the server. In this way we can interact with all the headset components.

\textit{ARVRServer} plays an important role because through it we can bind code to interact with \textit{SteamVR}. This is a virtual reality hardware and software that makes possible execute a virtual reality experience and communicate with the headset components. So ARVRServer represent a middle-layer between SteamVR and Godot. All available headset components that SteamVR detects are made available in Godot thanks to ARVRServer interfaces.

To interact with Arduino we use a plugin based on GDNative called \textit{GDSercomm}. It is a GDNative module that allows a serial port communication between Arduino and Godot.
Arduino is an open source hardware and software company, project and user community that designs and produces single-board micro-controllers kits for building digital devices. 
We used it to implement the vibration system under the wooden footplate.

To create the game environment like footplate, bridge and other objects with which the player can interact we use Blender. It is a free and open source 3D creation suite and supports the entire  3D pipeline – modeling, rigging, animation, simulation, rendering, compositing and motion tracking, even video editing and game creation. 


\section{The Virtual World made on Godot and Blender\label{virtualworld}}

To create a game with Godot the recommended approach is to think about the scenes that compose the project in a naturally way. So we started to imagine all the visible elements with which the player should interact, write down them in a diagram and creating for each one a Scene. 

\begin{figure}
    \centering
    \includegraphics[width=0.5\textwidth]{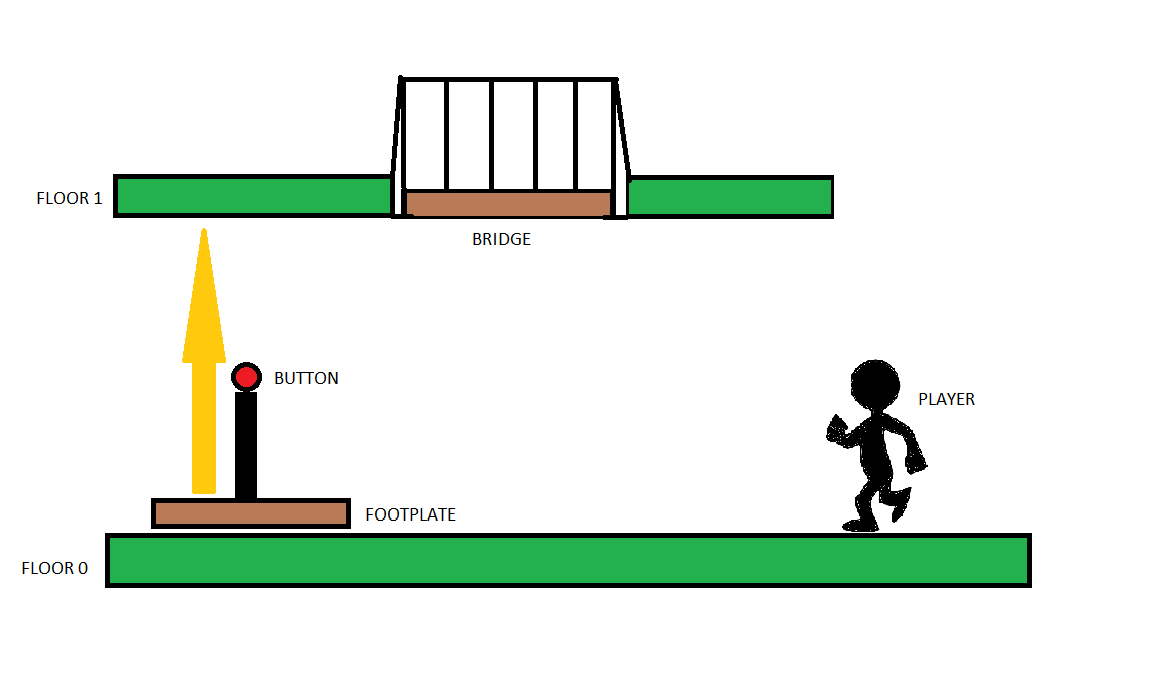}
    \label{fig:diagram}
    \caption{Diagram with all visible elements}
\end{figure}

Our final diagram is shown in the figure \ref{fig:diagram}.

The 4 basic elements that composed our game are:
\begin{itemize}
    \item \textbf{Environment}: composed by the first floor and second floor;
    \item \textbf{Player}: is composed by hands, feet, camera and an arrow that indicates the teleport end point;
    \item \textbf{Footplate} and \textbf{Button};
    \item \textbf{Tibetan bridge}.
\end{itemize}

Each of these elements has one or more Mesh assign to it. A Mesh is a collection of vertices, edges and faces that defines the shape of a polyhedral object.

To create the meshes we use Blender. It is a versatile and complete software, based on algorithms capable of reproducing the behavior of light in a realistic way. In addition to providing a great variety of features for modeling, lighting and rendering, the program allows a rather simple export of the models within a Godot Engine scene. In order to be displayed correctly, it is important to export the meshes created in Blender in one of the following ﬁles supported by Godot: \texttt{DAE}, \texttt{GLTF}, \texttt{OBJ} and \texttt{ESCN}. 

The first mesh that we create was the footplate, able to move the observer in the upper floor. This consists of a rectangular table to which we apply a wood texture. Blender provides options to improve some eﬀects related to opacity and color, once the object is hit by light. 
A similar method has been used for the construction of the Tibetan bridge because also in this case we started modeling from simple cubes and cylinders.

For the Player's hands and feet we download some free assets while, for the first and second floor, we use the basic mesh that Godot makes available.

Once all the meshes were exported in Godot we started to organized them in different scenes to better handle each game's component behavior. A \textbf{Scene} is a group of nodes organized hierarchically while a \textbf{Node} represents the core component of Godot and it can perform a variety of specialized functions.
Another important aspect that we have to take in mind in game development is the ability to detect some collisions through the objects; for this reason Godot provides four basic types of physical bodies able to detect and respond to collisions:
\begin{itemize}
    \item \textbf{Area}: provides detection and influence. It can detect when objects overlap and can emit signals when bodies enter or exit.
    \item \textbf{StaticBody}: is not moved by the physics engine, it participates to collision detection but does not move in response to the collision.
    \item \textbf{RigidBody}: implements simulated physics, we do not control it directly, but we apply forces to it (such as gravity, impulse, etc.) and the physical motor calculates the resulting movement. 
    \item \textbf{KinematicBody}: It is not influenced by physics. This type of physic body is useful for moving objects that do not require advanced physics. It must be controlled by the user.
\end{itemize}
To better understand these information we have to take in mind that each element that we mention before performs a specific role in the game. For this reason we have to use specific nodes in order to obtain the desired behavior.

For example, because the Environment is not intended to move, the elements that compose this scene are StaticBody. Each StaticBody wraps the mesh instance and the collision shape.

The Footplate is a KinematicBody because it has to move the player in the upper floor. This behavior is obtained using the  \texttt{move\_and\_slide} function that translates the footplate along the y axis and it stops once it reaches the upper floor. 
The Bridge is a RigidBody and we apply a rotation impulse to it when the Player walks over it.

The most complex Scene is represented by the Player. Basically it is a RigidBody but, because of it has to work in virtual reality environment, it has to implement some specifics nodes like: 
\begin{itemize}
    \item \textbf{ARVROrigin}: is a special node within the AR/VR system that maps the physical location of the center of our tracking space to the virtual location within our game world.
    \item \textbf{ARVRCamera}: is a camera node with a few overrules for AR/VR applied, such as location tracking.
    \item \textbf{ARVRController one for each hand and foot}: is a spatial node representing a spatially-tracked controller. In our project it wraps the hands' and feet's mesh istances and their collision shapes.
\end{itemize}
The structure of the Player is showed in the above image.
\begin{figure}
    \centering
    \includegraphics[width=0.3\textwidth]{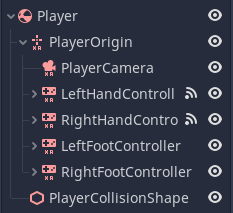}
    \caption{Player structure tree.}
    \label{fig:diagram2}
\end{figure}

Most of the elements that compose our scenes contains a \textbf{CollisionShape} node. It defines the object's collision bounds that is used to detect contact with other objects. In base of the shape of the object that it has to wrap we could use some basic types or create new complex collision shape  in Godot. It is important to know that is possible to create some complex collision shape inside other external software – like Blender – and import them in Godot.

One of the most powerful collision features that Godot provides is the collision layer system. It allows to build up complex interactions between a variety of objects. The key concepts are \textbf{layers} and \textbf{masks}.
The \textbf{collision\_layer} describes the layers that the object appears in. The \textbf{collision\_mask}  describes what layers the body will scan for collisions. Keeping track of what layer you are using could be difficult so Godot provides the possibility to rename the layer.
\\
In our project we have the following collision situation:\\
\includegraphics[width=0.3\textwidth]{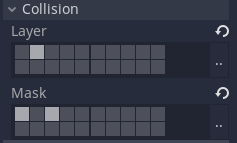}
\includegraphics[width=0.3\textwidth]{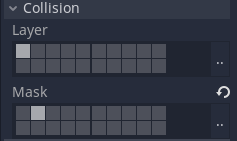}
\includegraphics[width=0.3\textwidth]{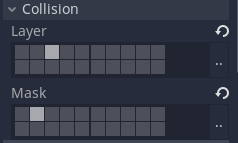}
\caption{1:Player collision 2:Footplate collision 3:Bridge collision}

The platform and the environment's floors appear in layer 1 and checks for collisions with player (layer 2).
The player appears in layer 2 and checks for collisions with environment (layer 1) and bridge (layer 3).
The bridge appears in layer 3 and checks for collisions with player (layer 2).

In order to obtain the desired behavior from Footplate, Bridge and Player we attach to them a \textbf{GDScript}. It is a high-level, dynamically typed programming language used to create content. It uses a syntax similar to Python (blocks are indent-based and many keywords are similar). Its goal is to be optimized for and tightly integrated with Godot Engine, allowing great flexibility for content creation and integration.

First of all we combined all the scenes created in the Main scene. We chose it as the main scene that Godot runs when it starts. Also the Main scene has a script attached to it because, when we start the game we have to check the availability of the ARVRServer and its devices. The following code shows how we check the state of ARVRServer and its interfaces.

\begin{lstlisting}[frame=single,caption={The code used to configure the Main scene.}]
extends Spatial

func _ready():
	var interface = ARVRServer.find_interface("OpenVR")
	if interface and interface.initialize():
		# turn to ARVR mode
		get_viewport().arvr = true
	
		# keep linear color space, not needed with the GLES2 renderer
		get_viewport().keep_3d_linear = true
	
		# make sure vsync is disabled or we'll be limited to 60fps
		OS.vsync_enabled = false
		
		# up our physics to 90fps to get in sync with our rendering
		Engine.target_fps = 90
\end{lstlisting}
From the code we can see that the Main scene is a Spatial node. Most of 3D game objects inherit from Spatial because it allows to move, scale, rotate and show/hide its children in a 3D project.

The function \texttt{\_ready()} is called only once when the node is created. In it is possible to initialize variables, load nodes, materials and all the things that we need to start a scene. 

In our case we try to find \textit{OpenVR} interface because, if it is available we initialize the virtual environment and sets the frame per second to 90 to avoid the user sickness.

Once all the interfaces are ready the game starts and all the scenes that we create are visible in the virtual world.

To handle the collision between the Player and the Footplate we attach to the Player the following script:
\begin{lstlisting}[frame=single,caption={The code used to configure the Player scene.}]
extends RigidBody

var collision
var new_position
var leftfoot
var rightfoot
var positionl
var positionr
var camera
var position
var changel = false
var changer = false

func _ready():
	leftfoot = get_node("PlayerOrigin/LeftFootController/LeftFootArea")
	positionl = leftfoot.get_translation()
	rightfoot = get_node("PlayerOrigin/RightFootController/RightFootArea")
	positionr = rightfoot.get_translation()
	camera = get_node("PlayerOrigin")
	position = camera.get_translation()
	collision = get_node("PlayerCollisionShape")
	new_position = collision.get_translation()
	
func _process(delta):
	positionl = leftfoot.get_translation()
	positionr = rightfoot.get_translation()
	position = camera.get_translation()
	if changel:
		new_position = collision.set_translation(position)
	elif changer:
		new_position = collision.set_translation(position)

func _on_LeftFootArea_body_entered(body):
	if body.get_name() == "BottomFloor" or body.get_name() == "UpperFloor1":
		if positionl != new_position or positionr != new_position:
			changel = true


func _on_RightFootArea_body_entered(body):
	if body.get_name() == "BottomFloor" or body.get_name() == "UpperFloor1":
		if positionl != new_position or positionr != new_position:
			changer = true


func _on_LeftFootArea_body_exited(body):
	if body.get_name() == "BottomFloor" or body.get_name() == "UpperFloor1":
		changel = false


func _on_RightFootArea_body_exited(body):
	if body.get_name() == "BottomFloor" or body.get_name() == "UpperFloor1":
		changer = false
\end{lstlisting}

First of all we initialize the controllers with which the Player will interact with the virtual world components. The hands can triggers – thank to the collision shape attached to them – if there are some objects that the Player can press or hold. The feet controllers are useful to handle the movements of the  Player. In fact, thank to these we can update the position of the Player in the virtual world every time he walks. 

The \texttt{\_process()} is a special function that updates object data every delta frames. We use it to update the position of the player every time he moved in the space.

The other functions that are in the script are used to handle specific signals emitted by some some Area nodes.

To handle the Player teleport we attach to LeftHandController node another script that create an arrow that became visible each time the trigger button of controller is clicked. This arrow show the end point to which the Player will be moved.

Very interesting is how we connect virtual Footplate with the real one. The connection between the Godot footplate node and Arduino was setup by a special GDNative module called \textbf{GDSercomm}. It is a module that allows a serial communication between Godot and Arduino, in other words it provides an API from them. It presents some methods like:
\begin{itemize}
    \item \textbf{list\_ports()}: to get all the available ports;
    \item \textbf{open()}: to open a communication, flush() - to update the buffer;
    \item \textbf{get\_available()}: return the available reading bytes;
    \item \textbf{write()}: write a string in the buffer.
\end{itemize}
All these function are contained in Sercomm, a C library.

\begin{lstlisting}[frame=single,caption={Some pieces of code used to configure the vibration motors from Godot.}]
func _ready():
	timer.connect("timeout", self, "_on_timer_timeout")
	timer_setup()
	upperFloor = get_node("../Environment/UpperFloor1")
	stopping = int(upperFloor.get_translation().y)
	print("ports ", PORT.list_ports())
	var ports = PORT.list_ports();
	PORT.open(ports[0], 9600, 1000)
	PORT.flush()
	print("get_available ", PORT.get_available())
	
[...] 

func _move_platform_with_button():
	player.set_translation(Vector3(1.7,0,0.8))
	origin.set_translation(Vector3(-1.7,0,-0.8))
	camera.set_translation(Vector3(-1.7,0,-0.8))
	timer.stop()
	is_platform_moving = true
	platform_moved = true
	area_mesh_instance.visible = false
	PORT.write("h")
	PORT.flush()
\end{lstlisting}
In the \texttt{\_ready()} function – in the last rows of code – we establish a connection with Arduino port.

The \texttt{\_move\_platform\_with\_button()} function is called when the Player press the button near the Footplate in virtual world. When it is pressed the player is moved upstairs and a string is send to Arduino. In this way we activate the vibration motors that are under the real footplate. To stop it we check if the virtual footplate reaches the upper floor and eventually we stop the translation in the y axis and send a string to Arduino to stop the vibration of the motors.

\begin{lstlisting}[frame=single,caption={The code used to stop the vibration motors from Godot.}]
func _physics_process(delta):  
	if is_platform_moving:
		# makes the platform move
		vel.y = force * delta
		move_and_slide(vel, pos)
		h = self.get_translation().y
		# platform stops
		if h >= stopping + 0.3:
			force = 0
			PORT.write("l")
			PORT.flush()
\end{lstlisting}
The \texttt{\_physic\_process()} is used when one needs a framerate-independent deltatime between frames. 

In the end, to handle the rotation of the Bridge when the Player walk over it, we use this script that checks when the \textbf{LeftFootArea} or \textbf{RightFootArea} collide with one bridge's board and when it occurs we apply an \texttt{apply\_torque\_impulse()} to each peace of Bridge the Player touches.
\begin{lstlisting}[frame=single,caption={The code used to move the bridge when the plyer walk over it.}]
func _on_Board_area_entered(area):
	if area.get_name() == "RightFootArea" or area.get_name() == "LeftFootArea":
		timer.start()
		$Boards.apply_torque_impulse(Vector3(0.02, 0.0, 0.0))
\end{lstlisting}
The final result that we reach is showed in the following image.

\cincludegraphics[width=1\textwidth]{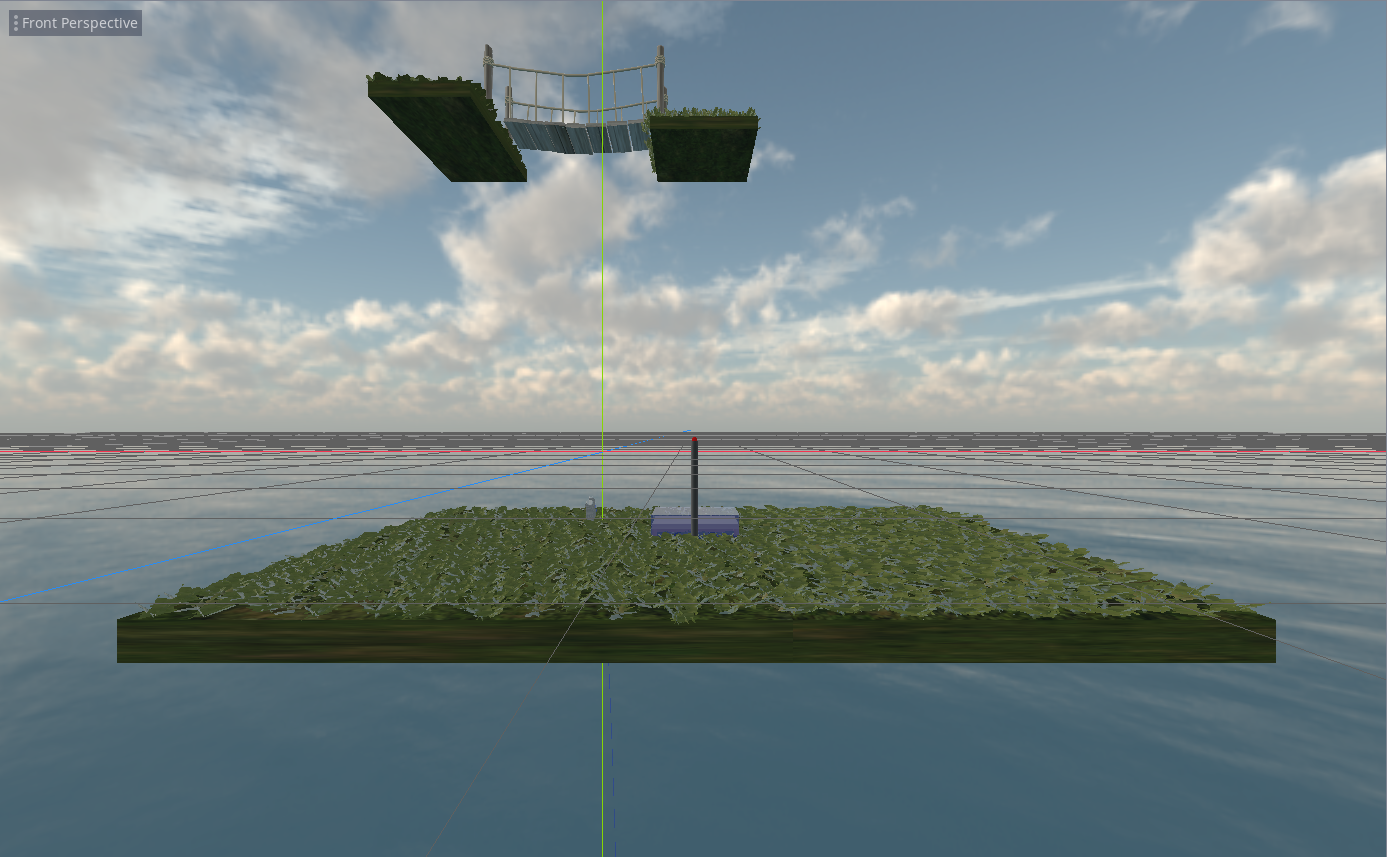}
\caption{Final result.}

\section{The vibration footplate and Arduino\label{}}

To make the user to feel the footplate moving under his feet we placed in the real environment a wooden footplate, hacked using an \texttt{Arduino Uno} board and two \texttt{vibration motors}.

Using the Arduino IDE we uploaded the code shown in listing \ref{listing:arduino} into the board which initialize and handle the serial communication.

In the setup function we initialize the serial communication at 9600 bps and wait until the serial port is ready to communicate, then we configure the pin 8 to behave as output.
The loop function does what the name suggests, loops consecutively and reads from the serial, when it receives the "h" character Arduino will set a 3.3 voltage to the pin 8 starting the vibration
motors, in contrary when it receives the "l" character Arduino set a 0 voltage to the pin 8 which stops the vibration motors.

\begin{lstlisting}[frame=single,caption={The code used to configure the Arduino board for vibrating the real footplate.},label=listing:arduino]

int incomingByte = 0;

void setup() {
  Serial.begin(9600);
  
  while(!Serial){
    ;
  }
  
  pinMode(8, OUTPUT);
}

void loop() {
  if(Serial.available() > 0) {
    incomingByte = Serial.read();
    
    if(incomingByte == 'h') {
        digitalWrite(8, HIGH);
    }
    else if (incomingByte == 'l') {
      digitalWrite(8, LOW);
    }
  }
}
\end{lstlisting}

\section{Conclusions and future works\label{conclusions}}
The project's purpose is to create an immersive VR experience from the design to the implementation. Through this project we tested the Godot's efficiency for VR games. We obtained good performances both in physics simulations and graphics quality.

In the beginning we created the virtual world space that has the same measurements as the real space in which the player moves, controlled by the HTC Vive system.
In this way the player can move in the virtual space without colliding with the room walls. 
In our case the simulation room was small so it was important to develop a circular and upwards path.

In the experimental phase we faced issues to setup the proper player interactions with other objects. A wrong layers collision setup could compromise the interaction behaviour of the objects or of the player.

The current project was presenting only a footplate as an interactive object. Our goal is to add more interactive objects, we could for example introduce a fan fixed on the ceiling. The fan will be activated with the vibration motors placed under the wooden footplate, to simulate the wind effect.

Furthermore the vibration system can be applied to other static objects.
%
%

\end{document}